\documentclass[10pt,prb,twocolumn,showpacs]{revtex4}
\usepackage{graphicx}
\usepackage{epsf}
\usepackage{ifthen}
\usepackage{amsmath}
\usepackage{amssymb}
\usepackage{subfigure}
\usepackage{psfrag}
\usepackage{float}
\usepackage{subeqnarray}
\begin{document}
\title{Rippling, buckling and melting of single- and multi-layer  MoS$_2$}

\author{Sandeep Kumar Singh$^1$, M. Neek-Amal$^1$, S. Costamagna$^{2,1}$,   and F. M. Peeters$^1$}

\affiliation{$^1$Universiteit Antwerpen, Department of Physics,
Groenenborgerlaan 171, BE-2020 Antwerpen, Belgium.\\ $^2$Facultad de
Ciencias Exactas Ingenier{\'\i}a y Agrimensura, Universidad Nacional
de Rosario and Instituto de F\'{\i}sica Rosario, Bv. 27 de Febrero
210 bis, 2000 Rosario, Argentina.}

\date{\today}

\begin{abstract}
Large-scale atomistic simulations using the reactive empirical bond
order force field approach is implemented to investigate thermal
and mechanical properties of single-layer (SL) and multi-layer (ML) molybdenum disulfide (MoS$_2$).
The amplitude of the intrinsic ripples of  SL-MoS$_2$ are found to be
smaller than those exhibited by graphene (GE).
Furthermore, because of the van der Waals interaction between layers,
the out-of-plane thermal fluctuations of ML-MoS$_2$ decreases rapidly with increasing number of layers.
This trend is confirmed by the buckling transition due to uniaxial stress which occurs
for a significantly larger applied tension as compared to graphene.
For SL-MoS$_2$, the melting temperature is estimated to be 3700~K which occurs through dimerization followed by
the formation of small molecules consisting of 2 to 5 atoms.
When different types of vacancies
are inserted in the SL-MoS$_2$ it results in a decrease of both
the melting temperature as well as the stiffness.

\end{abstract}
\pacs{68.60.Dv,62.20.-x}

\maketitle

\section{Introduction}

Two-dimensional (2D) transition-metal dichalcogenides (TMDCs)
have attracted a lot of attention due to the wide range of electronic
phases that they can exhibit, ranging from
metallic~\cite{Ivanovskaya2008, Lebegue2009, Kuc2011},
semiconductor~\cite{Ramasubramaniam, Topsakal, Ma} to
superconductor~\cite{Friend1987}. Recently, a lot of research efforts were devoted to
 MoS$_2$ due to its wide availability in nature as
molybdenite and its promising semiconducting characteristics (in
contrast to graphene which has a zero band gap). Bulk MoS$_2$ has an
indirect bandgap~\cite{Kam1982} whereas its single-layer has a direct
bandgap~\cite{Mak2010} and exhibits
photoluminescence~\cite{Splendiani2010} which is advantage for
optoelectronic applications. While it is known that the band gap can be tuned
by lattice deformations~\cite{nanolett2013}, the microscopic details of MoS$_2$ under
applied strain are still not well understood.

The phonon spectrum of MoS$_2$ is very different from that of graphene
resulting in distinct structural and thermal properties,
e.g. the well known negative thermal expansion
of graphene is not observed in MoS$_2$~\cite{cai}. There is also an energy gap
  of $\sim 50$~cm$^{-1}$ in the phonon spectrum of MoS$_2$ which separates
optical and acoustic phonon bands. The knowledge of the thermo-mechanical
properties of MoS$_2$ is crucial for the enhancement of the
performance of devices based on MoS$_2$. The role of defects on the
physical properties of monolayer MoS$_2$ is also important because
most of the 2D materials contain vacancies, which are
generated during the growth process~\cite{Najmaei2013, Zande2013} or
by ballistic displacements during imaging such as electron
irradiation, due to chemical etching and electron excitations  in high resolution transmission electron
microscopy~\cite{Meyer2009, Jin2009, Komsa2012, Gerardo2013, Recep2013}.
Recently, Zhou et al.~\cite{Nano_defect} found six types of point defects in monolayer
MoS$_2$ grown by chemical vapor deposition: i) monosulfur vacancy
($V_S$), ii) disulfur vacancy ($V_{S2}$), iii)
vacancy complex of Mo and nearby three sulfurs ($V_{MoS3}$), iv)
vacancy complex of Mo nearby three disulfur pairs ($V_{MoS6}$), and
v) antisite defects where a Mo atom substitutes a S$_2$ column
($Mo_{S2}$), or vi) a $S_2$ column substituting a Mo atom
($S2_{Mo}$).

\begin{figure}[t]
\includegraphics[width=0.5\textwidth]{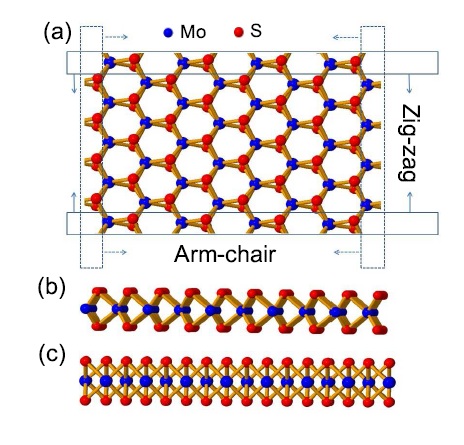}
\caption{ (Color online)  Schematic (a) top, (b) arm-chair and
(c) zig-zag side views of single layer MoS$_2$. Big-blue (small-red) circles refer to
the Mo (S) atoms. Dashed rectangles in (a) indicate the atoms that are fixed
during compression.
}\label{side_view}
\end{figure}

In our previous studies we investigated the thermo-mechanical properties of
different 2D-materials, e.g. graphene (GE), hexagonal boron-nitride (h-BN) and
their functionalized structures.~\cite{seba2012,sandeepBN2013,sandeepFG2013}
Here we report on the thermo-mechanical properties of single and multi-layer MoS$_2$.
Due to the S-Mo-S sandwich structure, we found exceptional mechanical stability and
a lower thermal excited rippling behavior as compared to graphene.
Similarly, the melting of MoS$_2$ occurs also at a lower temperature and exhibits
different microscopical characteristics such as the formation of small molecules.

The paper is organized as follows. Details of the MD simulation and a description of the modified
Lindemann parameter used to detect the melting transition are described in  Sec.~II.
Then, in Sec.~III we present results for the thermal ripples
and we compare them with those of GE.
In Sec.~IV we obtain the buckling transition for applied stress in the zig-zag and arm-chair directions.
We discus the melting behavior of MoS$_2$ together with the effect of
several kinds of vacancies on it in Sec.~V.
Finally, in Sec.~VI we present the conclusions of our work.


\section{Simulation method}
A proper interatomic potential function which is capable of describing accurately
the interactions in the material system is of crucial importance. Recently, a new approach based on bond-order
potentials emerged that depend on the local chemical environment in reactive simulations which capture
bond formation and breaking, saturated and unsaturated bonds, dangling and radical bonds, as well as single, double or triple bonds.
Liang et al.~\cite{Liang2009,Liang2012,Spearot2013} presented an interatomic potential for Mo-S systems which contains
a many-body reactive empirical bond-order (REBO) potential~\cite{Brenner} with a two-body Lennard-Jones (LJ)
potential. The REBO potential was chosen because it allows for bond breaking and bond formation during simulation.
The parameterized many-body Mo-S potential energy focuses primarily on the
structural and elastic properties of MoS$_2$ maintaining its transferability to other systems such as pure Mo
structures, low coordinated S, and some other binary structures.

The Mo-S many-body empirical potential has the following analytical form:
\begin{equation}
{E_b}=\frac{1}{2}\sum_{i} \sum_{j(>i)} [V^R(r_{ij})-b_{ij}V^A(r_{ij})].
\end{equation}

\noindent Here ${E_b}$ is the total binding energy, $V^R(r_{ij})$ and $V^A(r_{ij})$ are a repulsive and an attractive term,
respectively, with ${r_{ij}}$ the distance between atoms $i$ and $j$, given by

\begin{equation}
V^R(r)=f^C(r)(1+Q/r)Ae^{-\alpha r},
\end{equation}

\begin{equation}
V^A(r)=f^C(r)\sum_{n=1}^{3} B_n e^{-\beta r},
\end{equation}
where the cut-off function $f^C(r)$ is taken from the switching cutoff scheme.
The values for all the parameters used in our calculation for the Mo-S potential can be found in
Refs.~[\onlinecite{Liang2009,Liang2012}] and are therefore not listed here. Alternatively, it is also possible to use its competitor, the
Stillinger-Weber potential, to model the interaction between Mo-S, Mo-Mo and S-S~\cite{jiang}.



The mutual interaction between different S-Mo-S tri-layers is a
van der Waals (vdW) attraction between the S atoms which we
describe by the well-known Lennard-Jones potential,
 \begin{equation}
E_{LJ}(r)=4\epsilon\left[\left(\frac{\sigma}{r}\right)^{12}-\left(\frac{\sigma}{r}\right)^{6}\right],
\end{equation}
where r is the interatomic distance between S-S atoms,
$\sigma=3.3$\,\AA~and $\epsilon=6.93$\,meV.
The 12-6 Lennard-Jones potential parameters are used for the S-S
interaction such that the elastic constant c$_{33}$ of MoS$_2$ bulk is correctly reproduced~\cite{Liang2009}.

%

In the next section we study the thermal rippling, the mechanical properties
and the melting of a single- and bi-layer of MoS$_2$
using large scale atomistic simulations  with the above potentials. The Mo-S parameters
were implemented in the large-scale atomic/molecular massively parallel
simulator package LAMMPS~\cite{lammps,Plimpton}.

To account for the melting transition we analyzed the variation of
the total potential energy $E_T$ per atom with temperature
identifying partial contributions from Mo-atoms ($E_{Mo}$) and S-atoms
($E_S$). The Lindemann criterion~\cite{Lindemann} which states that the
system is melted when the root-mean-square (rms) value of the atomic
displacement is of the order of a tenth of the lattice constant, was used to characterize the
ordered state by considering the modified
parameter $\gamma$, used previously for 2D systems~\cite{Bedanov,Bedanov1,Zheng1}
and defined as
\begin{equation}
\gamma=\frac{1}{a^{2}}\langle| \textbf{r}_{i} -
\frac{1}{n}\sum_{j}\textbf{r}_{j}|^{2}\rangle,
\end{equation}
\noindent where a$=1/\sqrt{\pi\rho_{0}}$, $\rho_{0}$ is the 2D
particle density at T=$0$~K, $n$ is the number of nearest-neighbor
atoms, $\textbf{r}_{i}$ is the position of the $\textit{i}^{th}$
atom and the sum over $\textit{j}$ runs over the nearest-neighbor
atoms. Here, $i$ and $j$ were restricted to run only over Mo-atoms.


\section{Intrinsic ripples}

In order to study the thermal stability of MoS$_2$ we considered a square
shaped computational unit cell of MoS$_2$ (l$_x$=260~\AA~, l$_y$=280~\AA)
with both arm-chair and zig-zag edges in the $x$ and $y$ directions
with a total number of $N=25920$ atoms in the single layer
and $N=51840$ atoms in  bi-layer MoS$_2$.
In our simulation we adopted periodic boundary conditions and employed the NPT ensemble with $P$=0 using the
Nos\'{e}-Hoover thermostat and varied the temperature from 10\,K to 900\,K.

\begin{figure}[t]
\includegraphics[width=0.45\textwidth]{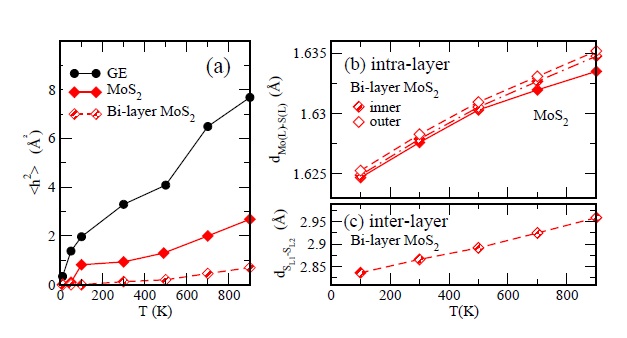}
\caption{(Color online) (a) Variation of $\langle h^2\rangle$
with temperature of  single layer MoS$_2$, bi-layer MoS$_2$ and graphene.
(b) Mo-S and (c) S-S bond lengths versus temperature.}\label{h2-ave}
\end{figure}

In Fig.~\ref{h2-ave}(a)
we show the evolution with temperature of
the average value of the height fluctuations $\langle h^2\rangle$
where $h$ refers to the height of the Mo atoms with respect to the center
of mass of the central Mo-layer. For comparison we have added here the
results of single layer graphene (circles) of comparable system size.
Notice that in the whole temperature range $\langle h^2\rangle$ for single layer MoS$_2$
is smaller than that of graphene. This result agrees with the estimated mechanical properties of these materials
where MoS$_2$ is expected to be more rigid~\cite{Liang2012}.
Notice that while the distance between the Mo atoms in the Mo-layer is larger than that of
the C atoms in graphene, it is the  Mo-S interaction that suppresses
the height fluctuations of the Mo atoms in MoS$_2$.

When a second layer of MoS$_2$ is added to form bi-layer MoS$_2$,
$\langle h^2\rangle$ is strongly reduced. This effect,
due to the S-S van der Waals type-interaction acting between the S-layers,
is fundamentally different from the Mo-S interaction.
The temperature-dependence of the intra-layer Mo-S and inter-layer
S-S (for bi-layer MoS$_2$) distances are shown in  Figs.~\ref{h2-ave}(b) and (c), respectively.
Notice that the distance Mo-S in bilayer MoS$_2$ is slightly larger than that of the single layer and
that there exists a small difference between the  S atoms from the
inner and the outer side which are more free to move.

The most adequate theory which allows to analyze in more depth  the
behavior of the intrinsic ripples is the elastic theory of continuum
membranes.~\cite{Nelson}
Its usage permits the detection of ripples with particular wave-lengths and also an
estimation of the anharmonic interactions in the
system. 
The key-quantity which
characterizes the behavior of the out-of-plane thermal fluctuations
is the height-height correlation function which in the harmonic limit has the following power law behaviour
$H(q)=\langle h(q)^2\rangle~\approx q^{-4}$.
%
The out-of-plane displacements of Mo atoms was analyzed by calculating $H(q)$ from
our molecular dynamics simulation by following the
same procedure as explained in our previous works~\cite{seba2012,sandeepBN2013}.

\begin{figure}[t]
\includegraphics[width=0.45\textwidth]{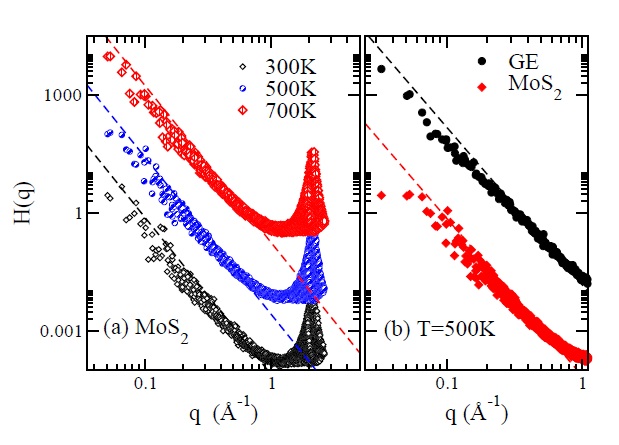}
\caption{(Color online) (a) Height-height
correlation function for MoS$_2$ at three different temperatures.
(b) Comparison of H(q) between GE and MoS$_2$ at 500~K.
}\label{hq}
\end{figure}

In Fig.~\ref{hq} we show $H(q)$ at 300 K, 500 K, and 700 K for
single layer MoS$_2$.
The results are shifted for a better comparison. The dashed-lines correspond to
the harmonic behavior and the peaked-structures at large wave-length are the
Bragg peaks of the crystalline lattice of the Mo-layer.
In all the cases $H(q)$ follows closely the $q^{-4}$ law.
In the long-wavelength regime, however, a larger fluctuation
of the points together with a deviation from the harmonic curve is observed.

In the case of GE, the stability of the membrane has been ascribed to the anharmonic coupling
between the in-plane and out-of-plane modes
which renormalizes the long wavelength ripples and prevents the occurrence of the crumpling
transition~\cite{fas1,dobry}.
In Fig.~\ref{hq}(b) we show $H(q)$ for GE together
with the one obtained for MoS$_2$, at 500~K.
While $H(q)$ in GE exhibits the expected deviation from the $q^{-4}$ harmonic
scaling due to the anharmonic interactions
at small $q$, in MoS$_2$ a larger fluctuation in the simulation results is present.
However, in the long-wave length regime a deviation from the harmonic law
still exist and it appears to be larger than that in GE.
This result is consistent with the lower value of $\langle h^2\rangle$ reported in Fig.~\ref{h2-ave}(a)
which is a consequence of the reduction in  long-wavelength ripples.

The origin of the breakdown of the harmonic behavior in MoS$_2$ is very different from the one in GE.
Because of the layered structure of MoS$_2$, its phonon modes and lattice vibration are
different from those in single layer graphene. The internal modes (due to the vibration of the Mo-S bonds)
are activated with lower energy with respect to e.g. the C-C bonds in graphene.
The latter is more susceptible to temperature
 making it a more floppy material.
This can be seen also from the results of next section where we investigate the buckling transition.
Therefore, we expect less coupling between out-of-plane and
in-plane modes in MoS$_2$  as compared to  graphene.


\section{Buckling transition}

\begin{figure*}
\includegraphics[width=0.32\textwidth]{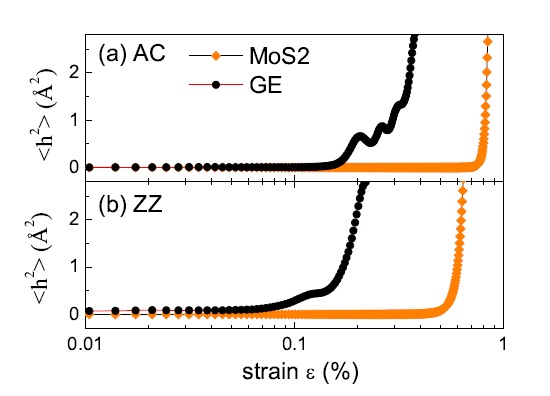}
\includegraphics[width=0.32\textwidth]{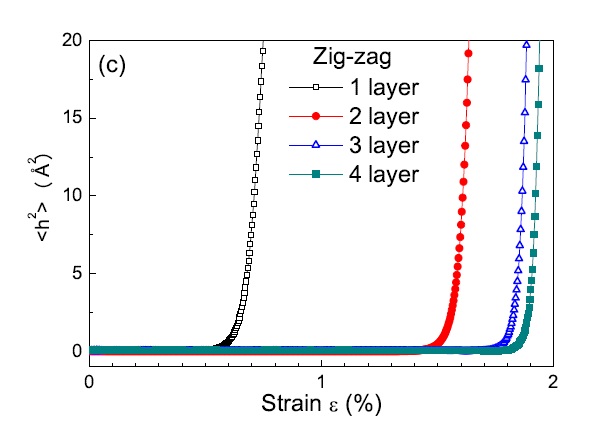}
\includegraphics[width=0.32\textwidth]{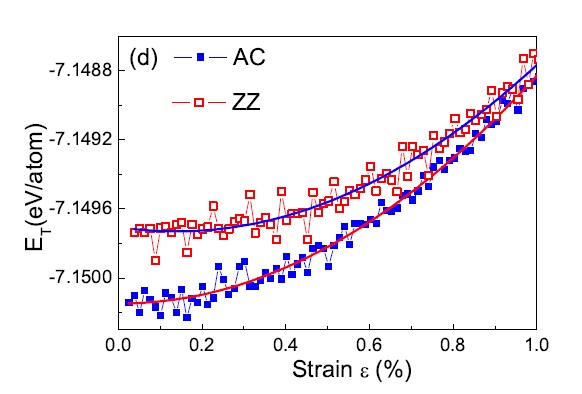}
\caption{(Color online) Variation of $\langle h^2\rangle$
in SL-MoS$_2$ with applied uniaxial stress along (a) arm-chair
and (b) zig-zag directions. For comparison purposes we show also
the corresponding results for graphene. (c) $\langle h^2\rangle$
of single-, bi-, tri- and four-layer MoS$_2$ versus uniaxial strain.
(d) Variation of the total energy with uniaxial strain for MoS$_2$ along arm-chair (blue) and zig-zag (red) directions.
}\label{buckling_1layer}
\end{figure*}

\begin{figure}
\includegraphics[width=0.48\textwidth]{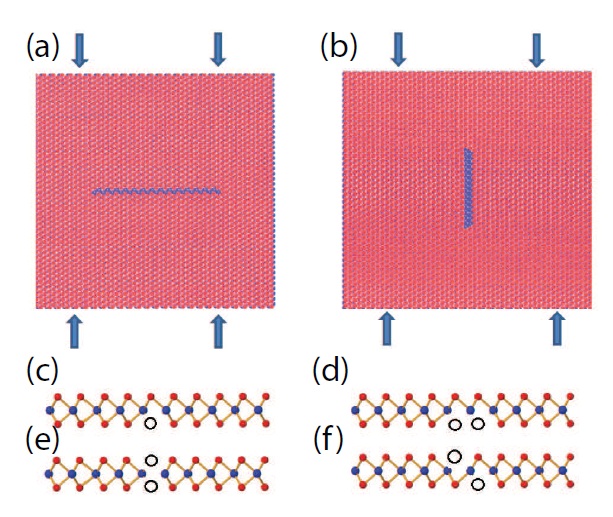}
\caption{(Color online) (a) perpendicular (b) parallel vacancy lines
with applied uniaxial stress along zig-zag direction. Various models for the
lines (c) single vacancy line, (d) two neighboring vacancy lines
in the same S layer (e) two vacancy lines coinciding in top and bottom layers,
and (f) two vacancy in staggered configuration.
}\label{buckling_linedefect}
\end{figure}

\begin{figure}
\includegraphics[width=0.48\textwidth]{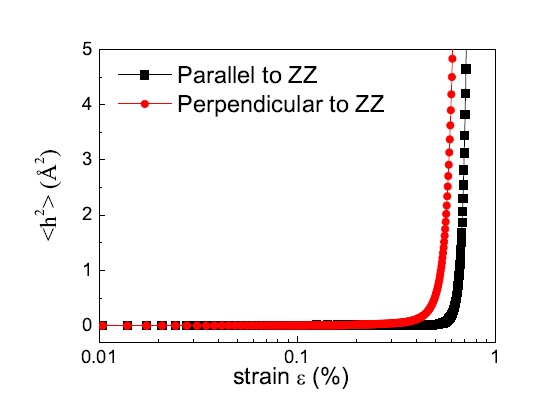}
\caption{(Color online) Variation of $\langle h^2\rangle$
in SL-MoS$_2$ for perpendicular and parallel vacancy lines with applied uniaxial stress along zig-zag direction.
}\label{buckling_line}
\end{figure}

The specific crystal structure of  MoS$_2$ influences its response to
external stress. Here we investigate the effect of  uniaxial compression stress
on the mechanical stability of MoS$_2$. We consider separately the zig-zag
and the arm-chair directions and fixed temperature to 10~K.
The outer row atoms are  fixed during the compression steps which
are indicated in Fig.~\ref{side_view} by the rectangular areas.
The compression rate was taken to be $\mu=0.5$\,m/s (for more details see
our previous studies~\cite{neekprb82,mehdi2}) which is small enough to guarantee
that the system is in equilibrium during the whole compression
process. 
The critical strain varies with applied compression rate and
the system size~\cite{jiang}. Recently Jiang et al.  studied the buckling of
single-layer MoS$_2$ under uniaxial compression using parameterized
Stillinger$–$Weber potential for MoS$_2$~\cite{Stlinger}. In this
section we restrict ourselves to those aspects that were not investigated in Ref.~[\onlinecite{jiang}].

Figure~\ref{buckling_1layer}(a) shows the variation of
$\langle h^2\rangle$ with applied uniaxial strain which was
determined using  $\epsilon=\mu t/l$ where $t$ is the time (after
starting the compression) and $l$ is the initial length in the
direction of the compression. The buckling transition occurs for
$0.60\%$ strain, which is about seven times larger than  GE ($0.09\%$) when the stress
is applied along the zig-zag direction and  $0.80\%$ which is about five times larger than the
one for GE ($0.16\%$) when  uniaxial stress is applied along the
arm-chair direction.
Notice that the buckling transition in MoS$_2$ is
sharper than in GE which is attributed to the sandwich structure of MoS$_2$.

The different responses of multilayer MoS$_2$ on the applied uniaxial stress
are displayed in Fig.~\ref{buckling_1layer}(c).
Here MoS$_2$ flakes with dimension $l_x\times l_y=14\times 14$\,nm$^2$ are considered.
It is clear that the single layer becomes buckled at smaller strains as compared
to the cases of bi-, tri- and four-layer  for which the buckling transition takes place at
$1.5\%$, $1.8\%$ and $1.85\%$, respectively.
In particular, tri- and four-layer are close to each other and therefore they approach
the limit of bulk MoS$_2$.

Uniaxial stress simulations can also be used to estimate the Young's modulus.
The results for applied stress in the longitudinal (arm-chair direction)
and lateral (zig-zag direction) directions are shown in Fig.~\ref{buckling_1layer}(d).
The Young's modulus is found by fitting the total energy (per area) to the quadratic function:
\begin{equation}
E_T=E_0+\frac{1}{2}Y\epsilon^2, \label{Eeff}
\end{equation}
where $Y$ is the Young's modulus of the system. Using aforementioned
fitting process, $Y$ is calculated  for a flake with arm-chair and
zig-zag MoS$_2$, to be 149 N/m and  148 N/m respectively, which are values between the recent DFT result~\cite{Cooper,Cooper2013} 130
N/m and the experimental value 180$\pm$60
N/m~\cite{Bertolazzi}.
Notice that the Young's modulus of graphene is  340 N/m which is 2.25 times larger than that of MoS$_2$.

Notice that since graphene is one-atomic-thick
structure it is extremely soft in the out-of-plane direction. The latter results in
much lower bending modulus than MoS$_2$~\cite{jiang_mos2}.  This is the reason
for higher $<h^2>$ in Fig.~\ref{h2-ave}(a) for graphene with respect to MoS$_2$.
However the in-plane stiffness of graphene because of strong in-plane sp$^2$ bonds
is expected to be much larger than MoS$_2$.

We also studied the effect of vacancies on the buckling transition. Notice that vacancies
alter the structure of MoS$_2$ and change the internal bonds between atoms. This results in a
change of the response of the system to any external force simply because the stiffness of
the system is reduced even for a few vacancies.~\cite{Neek-amal235437,Neek-amal153118}
Recently Komsa et al.~\cite{Komsa2013} studied sulfur vacancies in monolayer MoS$_2$ under electron irradation
using high-resolution transmission electron microscopy. These single vacancies are
mobile under the electron beam and tends to agglomerate into lines, where the direction of line defects
is sensitive to applied uniaxial stress. Figs.~\ref{buckling_linedefect}(a,b) present the
vacancy lines perpendicular and parallel to the applied uniaxial stress, respectively. The
buckling transition  for staggered double vacancy lines (which are more favorable in experiments)
perpendicular and parallel to the applied stress occurs for $0.4\%$
and $0.6\%$ strain (see Fig.~\ref{buckling_line}) where later is close to the pristine system.

\section{Melting behavior}

We investigate now the microscopical characteristics of the melting process of a single layer MoS$_2$.
Due to the large simulation time for each temperature
we considered here a smaller square shaped computational unit cell having $N$=7290
(Mo=2430 and S=4860) atoms. The simulations were performed in  the NPT (P=0) ensemble with periodic
boundary conditions.  Temperature was maintained by the
Nos\'{e}-Hoover thermostat and the MD time-step was taken to be
0.1~fs.

\begin{figure}
\includegraphics[width=0.50\textwidth]{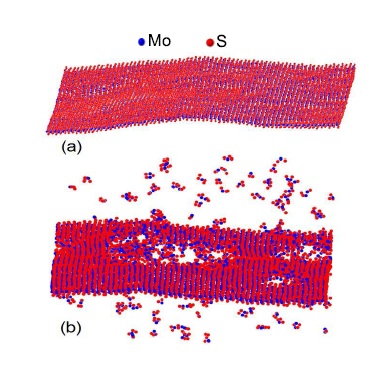}
\includegraphics[width=0.35\textwidth]{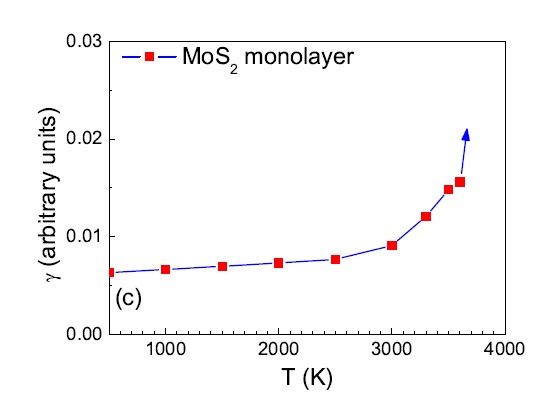}
\vspace{-0.6cm}\caption{(Color online) Two snap shots of our MD
simulation for the melting of MoS$_2$ which were taken (a) before
melting at 3600 K, and (b) during the melting at
3700\,K. (c) Modified Lindemann parameter $\gamma$ versus temperature.}\label{melting}
\end{figure}

\begin{figure}
\includegraphics[width=0.45\textwidth]{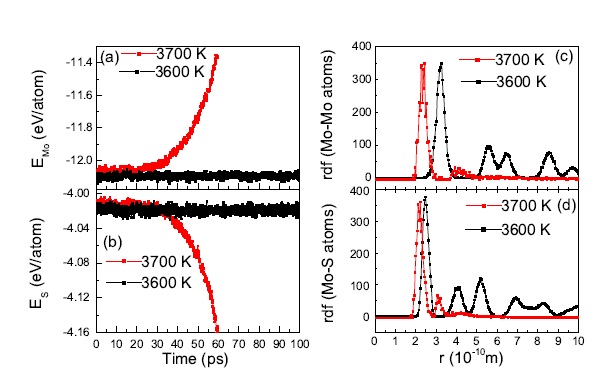}
\vspace{-0.6cm}\caption{(Color online) The variation of the potential
energy with time of (a) Mo and (b) S atoms before and after melting. In (c)
and (d) we show the variation of the radial distribution function of
Mo-Mo atoms and Mo-S atoms, respectively. }\label{melting_1layer}
\end{figure}

We first analyze the case of single layer MoS$_2$ and keep  separately track of the Mo and S potential energy contribution.
In Fig.~\ref{melting}, we show two snap shots of the system before (a) and during (b) melting.
The melting temperature T$_m$=3700 K is confirmed by the Lindemann parameter $\gamma$ (Fig.~\ref{melting}(c))
for only the Mo atoms and their nearest neighbors. $\gamma$ increases linearly as temperature
increases and diverges at melting.
Figures~\ref{melting_1layer}(a,b) show the variation of the potential energy per atom with time for Mo and S
atoms, i.e. $E_{Mo}$ and $E_S$ respectively, at $3600$~K and $3700$~K.
The sharp increase (decrease) in $E_{Mo}$ ($E_{S}$) is a signature of melting at T$_m\sim3700$~K.
After melting,  the Mo-atoms remain bonded to the S-atoms and form small molecules which is the
reason for the observed increase and decrease of the energy in Figs.~\ref{melting_1layer}(a,b).
The larger reduction in $E_S$ indicates that the S atoms prefer to be bonded
to the Mo atoms rather than to result in free S atoms.

The radial distribution function (rdf) of Mo-Mo shows that
before melting there is a sharp peak around 3.2~\AA~ that after melting is shifted to the range 2.1\AA- 2.5\AA~ which
is the distance between Mo-atoms in small MoS clusters (see
Fig.~\ref{melting_1layer}(c)).
However, the Mo-S rdf in Fig.~\ref{melting_1layer}(d) shows that after melting there are only
two peaks around 2.2~\AA~and 3.2~\AA~which are due to the formation of Mo-S and Mo-S$_2$ clusters.
In contrast to graphene where after melting the sample turns into random chains of carbon~\cite{Zakhar,sandeepFG2014},
SL-MoS$_2$ transits to a  phase consisting of MoS$_X$ clusters.
Thus at melting, atoms fluctuate around their equilibrium
position, the inter-atomic Mo-Mo bonds are broken and Mo becomes free and forms
clusters with S atoms.

\begin{figure}
\includegraphics[width=0.50\textwidth]{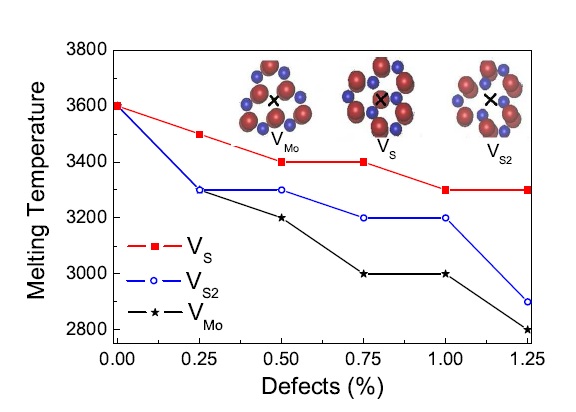}
\caption{(Color online) Melting temperature of MoS$_2$ as function of the percentage of vacancies for
different vacancy defects namely $V_{S2}$, $V_{MoS3}$ and $V_{MoS6}$. The cross symbols in the insets
indicate the missing atom in each structure.}\label{melting_defect}
\end{figure}

Previously it was found in the presence of helium gas at 1
bar pressure that the melting of bulk MoS$_2$  occurred around
1853-1895~K depending on the rate of
heating~\cite{Vasilyeva}. It was found that at high heating rate
MoS$_2$ began to decompose into MoS$_3$ and sulfur gas starting from
the solid phase. At lower rates the evaporation losses increased
markedly and MoS$_2$ was converted into the solid Mo$_2$S$_3$ and Mo
gas which is a mixture composed of variable amounts of the phases
identified chemically and structurally. It was shown that
the helium gas pressure had an influence on the melting temperature. Our results
show almost a factor of two higher melting temperature. The reasons
for this substantial larger melting temperature may be the
presence of helium gas in the experiment, the presence of defects (see
 Fig.~\ref{melting_defect}), dislocations in experimental
sample, and the weak vdW interaction between layers. Recently, using high resolution
electron microscopy imaging the atomic structure and morphology of grain boundaries
in MoS$_2$ have been reported~\cite{Sina}. As we show in Fig.~\ref{melting_defect}
any kind of vacancy in the system reduces the melting temperature. In multilayer MoS$_2$
the melting starts
 at the outer layer also known as "surface melting", while in
single layer the melting occurs when the bonds between Mo and S are
broken. We also calculated the melting of bi-layer and tri-layer
MoS$_2$ and found a similar melting temperature as for single layer
MoS$_2$. Nevertheless, the melting problem of MoS$_2$ should
be studied systematically by performing tests using different
potentials as well as studying the size effect. We emphasize that using
any bond order type potential, the same order of melting
temperature is found which is not expected to be responsible for the factor of two difference
in melting temperature between bulk and SL-MoS$_2$. It is also important to note that
the time scale in our MD simulation is restricted to maximum nano-seconds which
will lead to an overestimation of the melting temperature, while a real
melting phenomena occurs in seconds.

It is worthwhile to study the effect of atomic
vacancies in MoS$_2$ on the melting temperature. We performed
several  simulations for MoS$_2$ with different percentage of vacancies (Mo, S,
S2)~\cite{Nano_defect} randomly distributed along the sample.
The presence of atomic defects in the MoS$_2$ sheet makes it less
stiff and consequently results in a lowering of the melting
temperature as can be observed in Fig.~\ref{melting_defect}.
It is clear that monosulfur vacancies (V$_S$), which is usually observed in experiments
due to its lower formation energy,  has little impact on the melting in
comparison to disulfur vacancies (V$_{S2}$) and V$_{Mo}$. We can
conclude that the presence of V$_{Mo}$ type defects makes MoS$_2$
thermally more unstable than the other  type of vacancies. Therefore missing either Mo
or S atoms reduces the melting temperature significantly. This may only indicate that the
experimental sample in Ref.~\onlinecite{Vasilyeva} is not perfect.

\section{Conclusion}
Different thermal and mechanical properties of multi-layer MoS$_2$ were investigated
using atomistic simulations. The melting temperature of MoS$_2$
was found to be very weakly dependent of the number of layers and is lower than the one
for graphene. MoS$_2$ transits quickly to a phase with MoS$_X$ clusters without
 the appearance of random coils unlike graphene and
graphite. The buckling transition in MoS$_2$ under uniaxial
compression is independent of the direction of the applied stress
which is also different from graphene. We found that the sandwich
structure of MoS$_2$ makes it a less stiff material with respect to
graphene and it was found to affect different physical properties.
MoS$_2$ is more sensitive to temperature  and less energy is needed
to excite vibrational modes. The buckling transition is sharper as
compared to that of graphene and occurs at substantially larger
values of strain. We found that perfect MoS$_2$ has a  higher melting
temperature than those systems with defects. The melting temperature
of  MoS$_2$, and MoS$_2$ with grain boundaries demand more theoretical and experimental studies using
different sizes of computational unit cells and very long MD simulation time which are beyond the aim of the present study.

\section{ACKNOWLEDGMENTS} This work is
supported  by the ESF-Eurographene project CONGRAN, the Flemish
Science Foundation (FWO-Vl) and the Methusalem Foundation of the
Flemish Government. We acknowledge funding from the
FWO~(Belgium)-MINCyT~(Argentina) collaborative research project.
We would like to acknowledge Douglas E. Spearot~\cite{Spearot2013} for giving
us the implemented parameters of MoS$_2$ in LAMMPS.

\end{document}